\documentclass[aps,prb,epsfig,showpacs,superscriptaddress,twocolumn]{revtex4-1}
\usepackage{graphicx}
\usepackage{geometry}
\usepackage{dcolumn}
\usepackage{bm}
\usepackage{csquotes}
\usepackage{color}
\makeatletter

\newcommand{\Rmnum}[1]{\expandafter\@slowromancap\romannumeral #1@}
\makeatother

\begin{document}
\title{\bf Effect of A-site ionic radius on metamagnetic transition in charge ordered $Sm_{0.5}(Ca_{0.5-y}Sr_{y})MnO_3$
  compounds}
\author{Sanjib Banik}
\email{sanjib.banik@saha.ac.in}
\affiliation{CMP Division, Saha Institute of Nuclear Physics, HBNI, 1/AF-Bidhannagar, Kolkata 700 064, India}
\author{Kalpataru Pradhan}
\email{kalpataru.pradhan@saha.ac.in}
\affiliation{CMP Division, Saha Institute of Nuclear Physics, HBNI, 1/AF-Bidhannagar, Kolkata 700 064, India}
\author{I. Das}
\email{indranil.das@saha.ac.in}
\affiliation{CMP Division, Saha Institute of Nuclear Physics, HBNI, 1/AF-Bidhannagar, Kolkata 700 064, India}

\begin{abstract}
  We investigate the ultra-sharp jump in the isothermal magnetization and the resistivity
  in the polycrystalline $Sm_{0.5}(Ca_{0.5-y}Sr_{y})MnO_3$ $(y = 0, 0.1, 0.2, 0.25, 0.3, 0.5)$
  compounds. The critical field $(H_{cr})$, required for the ultra-sharp jump, decreases
  with increase of `Sr' concentration, i.e. with increase of average A-site ionic radius
  $\langle r_A\rangle$. The magnetotransport data indicate that the phase separation
  increases with the increase of $\langle r_A\rangle$, i.e. with $y$. The dependency of
  $H_{cr}$ with magnetic field sweep rate reveals that the ultra-sharp jump from
  antiferromagnetic (AFM) state to the ferromagnetic (FM) state is of martensitic in nature.
  Our two-band double exchange model Hamiltonian calculations show that the 
  `Sr' doping induces the ferromagnetic clusters in the antiferromagnetic insulating phase
  and in turn reduces the critical field. In the end we present a phenomenological picture
  obtained from our combined experimental and theoretical study.
\end{abstract}

\maketitle

\section{Introduction}

Recently, materials exhibiting field induced metamagnetic phase transition between
two energetically competing phases have attracted a lot of attention due to its complex
nature~\cite{Mahendiran,Autret,Fisher,Hardy1,KrishnaM}. The occurrence of metamagnetic
transition is perceptible by the sharp jump in isothermal magnetization. It is well
established that this transition is independent to the microstructure and actually
related to the intrinsic nature of the materials\~cite{Ouyang}. Many extensive studies
on this magnetic field induced metamagnetic transition have been carried out over the
last decades~\cite{Roy,Velez,Choi,Danjoh}. Examples of such materials studied include
certain phase separated manganites, inter-metallic alloys such as $Nd_5Ge_3$, $Gd_5Ge_4$,
$CeF_2$, etc. and some phase separated well known multiferroic $Eu_{1-x}Y_xMnO_3$ systems.
The appearance of the magnetization steps is found to be sensitive to the cooling
magnetic field as well as on the magnetic field sweep rate~\cite{Mahendiran,Hardy1}.
Hardy et al~\cite{Hardy2} have shown that, in $Pr_{0.5}Ca_{0.5}Mn_{1-x}Ga_{x}O_3$, the
spontaneous magnetization jump occurs in the time evolution of magnetization
for a fixed temperature and magnetic field. Wu et al.~\cite{Wu} have also observed the
same phenomenon in manganite thin films. The observation of magnetization (and resistivity)
steps in $(La_{0.5}Nd_{0.5})_{1.2}Sr_{1.8}Mn_2O_7$ is also reported by Liao
et al~\cite{Liao}. But, the origin of these metamagnetic transition is still a matter of
investigation. Very different kind of mechanisms have been proposed, such as field
dependent orbital ordering in $Pr_{0.5}Ca_{0.5}Mn_{0.95}Co_{0.05}O_3$~\cite{Mahendiran},
spin quantum transition in $Pr_{5/8}Ca_{3/8}MnO_3$~\cite{Cao}, spin reorientation in
$FeRh$ thin films~\cite{Bordel}, geometrical frustration in garnets~\cite{Tsui}, spin
flop transition in $Ca_3CoMnO_6$~\cite{Flint} and burst like growth of the ferromagnetic
fraction in the phase separation picture. According to the most of the authors the origin
of the magnetization steps is of martensitic in nature~\cite{Hardy1,Hardy2,Hardy3}. In
spite of having lot of study to analyze metamagnetism in various systems the origin through
detailed analysis has been rarely addressed.

In manganites the metamagnetic transition is usually observed in low bandwidth charge
ordered systems. Therefore, in our investigation, to understand the origin of metamagnetic
transition we have chosen $Sm_{0.5}Ca_{0.5}MnO_3$ (SCMO) as parent compound, which is one
of the lowest bandwidth and robust charge ordered system. It needs 470 kOe magnetic field
at 4 K for the metamagnetic transition~\cite{Tokura}. As the electronic bandwidth depends
on the average A-site ionic radius $\langle r_A\rangle$~\cite{Moritomo,Mathieu}, our study
by changing the average A-site ionic radius will give us the lead to figure out the origin
of the metamagnetic transition. To obtain materials with different electronic bandwidth,
we replace `Ca' ions in $Sm_{0.5}Ca_{0.5}MnO_3$ compound by `Sr' ions. This `Sr' doping
undoubtedly increases the $e_g$ electronic bandwidth as `Sr' has higher ionic radius as
compared with `Ca'. By varying the concentration of 'Sr' doping we prepared a set of samples
with different $\langle r_A\rangle$ and analyzed their structural, magnetic and electrical
transport properties. We observe that the critical field decreases with the increase of
$\langle r_A\rangle$. We explain this by taking the induced ferromagnetic clusters with
'Sr' doping in to account, which act as the nucleation centers for the metamagnetism. In
addition, we performed spin-fermion Monte Carlo calculations using double exchange model
Hamiltonian to support our experimental results.

\begin{figure*}
\includegraphics[width=0.95\textwidth]{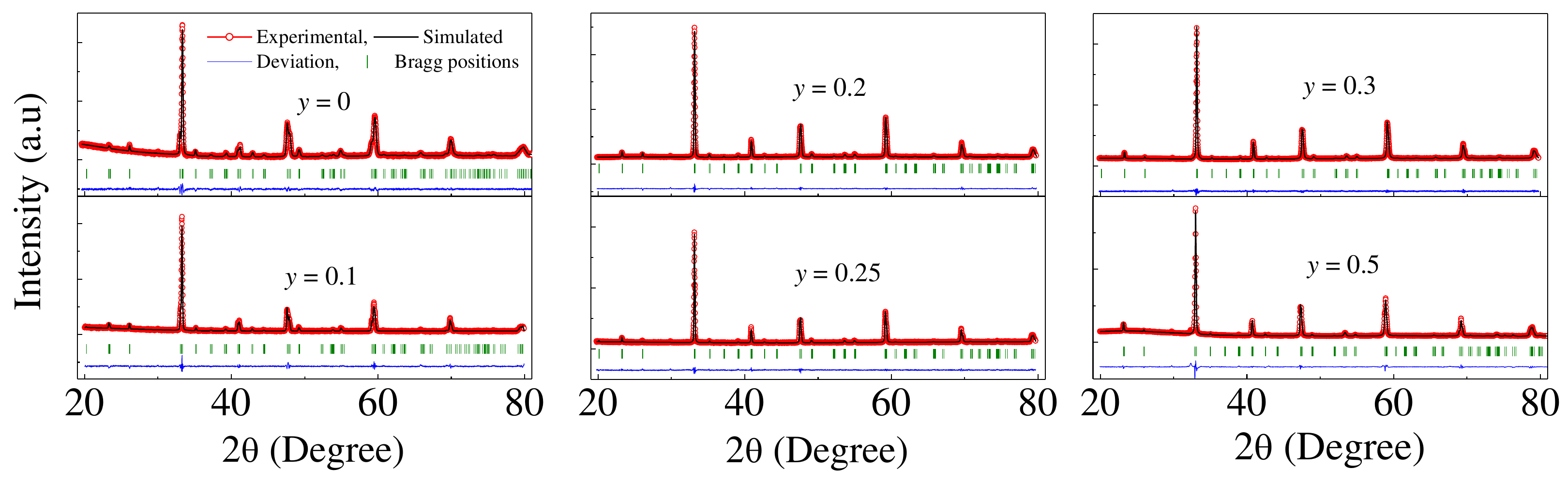}
\centering \caption[ ]
{\label{x-ray} Room temperature XRD data with its corresponding profile fitted data
for the compounds $Sm_{0.5}(Ca_{0.5-y}Sr_{y})MnO_3$ (${y=0,0.1,0.2,0.25,0.3,0.5}$)}
\end{figure*}

\section{Sample preparation and characterization}

All the bulk polycrystalline compounds $Sm_{0.5}(Ca_{0.5-y}Sr_{y})MnO_3$ (${y = 0-0.5}$)
have been prepared by the well known  sol-gel method with $Sm_2O_3$, $CaCO_3$, $Sr(NO_3)_2$
and $MnO_2$ as the starting chemicals of purity $99.9\%$. In order to prepare the bulk
samples decomposed gel has been pelletized and heated at $1300^0C$ for 36 hours.

The single phase nature of the samples have been characterized from room temperature
x-ray diffraction (XRD) measurements by using Rigaku-TTRAX-III with 9 kW rotating
anode copper source of wavelength $\lambda=1.54\AA$. Magnetic measurements have been
performed using quantum design SQUID-VSM. The transport and magnetotransport
measurements have been carried out on bar shaped samples in longitudinal geometry
by four probe method using Cryogenic setup.

\section{Results and discussion}

\subsection{Structural characterization}
The room temperature XRD study (Fig.~\ref{x-ray}) display the single phase nature of
all the bulk polycrystalline compounds. The crystal structure information has been
obtained from Rietveld refinement of the XRD data using FULLPROF software which shows
that all the samples crystallize in orthorhombic structure with `Pnma' space group.
The extracted lattice parameter and the average A-site ionic radius $\langle r_A\rangle$
are calculated from shanon effective ionic radii for different $y$
(see Table.~\ref{LATTICE_PARAMETERS}). $\langle r_A\rangle$ increases gradually  with
`Sr' concentrations because of its larger ionic radius as compared with `Ca'.

\begin{table}[h!]
  \caption{The lattice parameters and average A-site ionic radii for the samples
    $Sm_{0.5}(Ca_{0.5-y}Sr_{y})MnO_3$ (${y=0,0.1,0.2,0.25,0.3,0.5}$) }
\centering
\begin{tabular}{@{\hskip 0.2in}c   @{\hskip 0.2in}c   @{\hskip 0.2in}c   @{\hskip 0.2in}c   @{\hskip 0.2in}c}                        
\hline\hline                                                                                                                         
\textit{y}  &   a ($\AA$)    &   b ($\AA$)   &   c ($\AA$)   &   $\langle r_A\rangle$ ($\AA$)               \\ [0.5ex]               
\hline                                                                                                                               
0           &   5.423   &   5.370    &  7.582   &     1.156                                                 \\                      
0.1         &   5.415   &   5.381    &  7.593   &     1.169                                                 \\                      
0.2         &   5.410   &   5.401    &  7.626   &     1.182                                                 \\                      
0.25        &   5.404   &   5.413    &  7.629   &     1.188                                                 \\                      
0.3         &   5.410   &   5.416    &  7.634   &     1.195                                                 \\                      
0.5         &   5.441   &   5.425    &  7.660   &     1.221                                                 \\ [1ex]                
\hline
\end{tabular}
\label{LATTICE_PARAMETERS}
\end{table}

\begin{figure}[h!]
\includegraphics[width=0.48\textwidth]{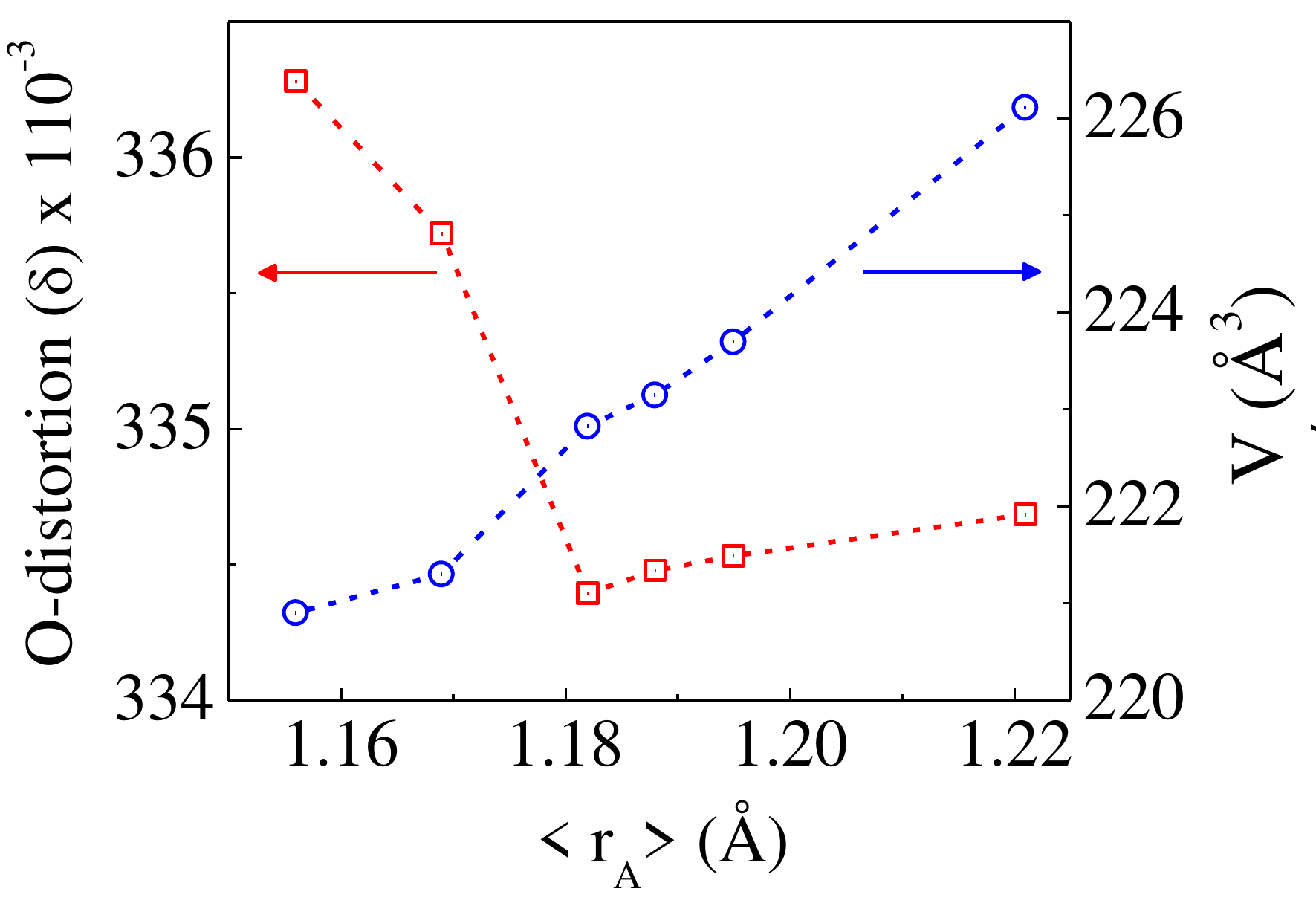}
\centering \caption[ ]
{Evolution of orthorhombic distortion ($\delta$) and unit cell volume with $\langle r_A\rangle$.}
\label{DISTORTION}
\end{figure}

We estimate the orthorhombic distortion [defined as
  ${\delta= \frac{a + b - c/\sqrt{2}}{a + b + c/\sqrt{2}}}$] from the lattice
parameters. The variation of $\delta$ and the unit cell volume with $\langle r_A\rangle$
are plotted in Fig.~\ref{DISTORTION}. Initially distortion decreases rapidly up to $y = 0.2$
with increase of $\langle r_A\rangle$ and then increases very slowly with $\langle r_A\rangle$.
On the other hand, unit cell volume increases steadily with increase of $\langle r_A\rangle$.

\subsection{Magnetotransport and magnetization study}

\begin{figure}[h!]
\includegraphics[width=0.48\textwidth]{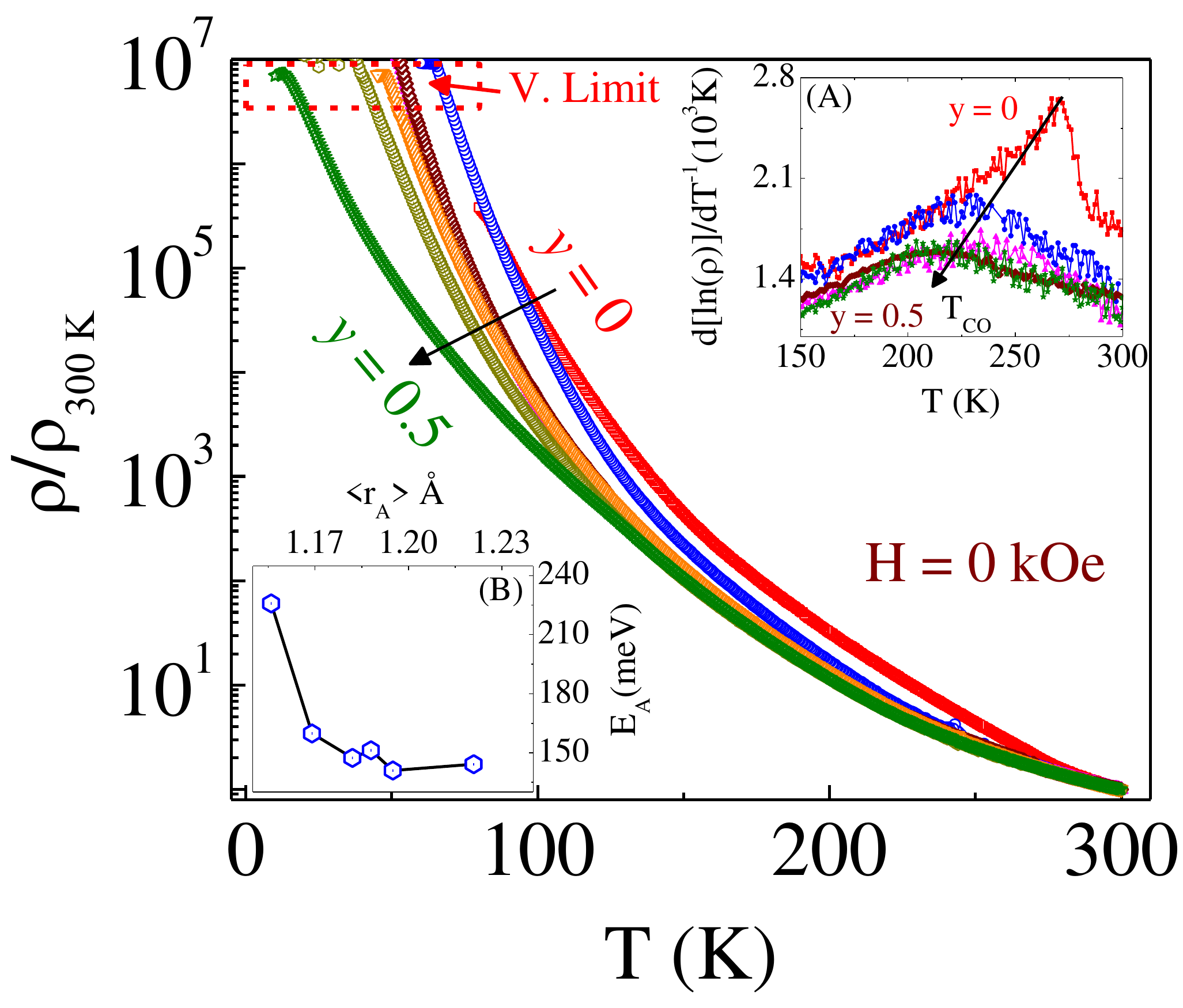}
\centering \caption[ ]
{\label{RT0} Temperature dependence of reduced resistivity of the samples
$Sm_{0.5}(Ca_{0.5-y}Sr_{y})MnO_3$ (\textit{y} = 0, 0.1, 0.2, 0.25, 0.5) in
absence of external magnetic field. Inset (A) shows the $d[ln(\rho)]/d T^{-1}$
vs. T plot for the samples and inset (B) shows the variation of activation energy
with $\langle r_A\rangle$.}
\end{figure}

The increase of average A-site ionic radius and reduction of orthorhombic distortion
greatly influences the transport and magnetotransport properties as the bandwidth of
$e_g$ electrons in manganites is directly proportional to the $\langle r_A\rangle$.
The temperature dependence of resistivity ${[\rho(T)]}$ in absence of any external
magnetic field has been performed for all the samples. The measurements were done
during warming cycle after cooling the samples in zero magnetic field. The evolution
of ${\rho(T)}$ with different `Sr' concentrations ($y$) (Fig.~\ref{RT0}) shows that
the samples are insulating down to the measurable resistance limit.

There is a relative suppression of resistivity value with increase of `Sr' concentrations
at low temperatures. Moreover, the `Sr' substitution also decreases the charge ordering
temperature ($T_{CO}$) from 260 K for ${y = 0}$ to 210 K for ${y = 0.5}$ [see inset (A) of
  Fig.~\ref{RT0}] and increases the fragility of the CO state. The reduction of resistivity
with `\textit{y}' as well as softening of CO state is due to the increase of bandwidth by
increasing A-site ionic radius. For further investigation, we analyze the high temperature
(${T > T_{CO}}$) resistivity data with the help of small polaron hopping
model (SPH). It is known that in manganites the electrical resistivity in paramagnetic
region is mainly governed by the polaronic activation. According to the SPH model~\cite{Banik}
the expression of resistivity is $\rho = \rho_0 T exp(E_A/k_B T)$ where $E_A$ is the polaronic
activation energy. From the fitting of the high temperature (${T > 260 K}$) resistivity data,
the activation energy $E_A$ for `Sr' doped samples has been calculated and its
evolution with `\textit{y}' is shown in the inset (B) of Fig.~\ref{RT0}. The activation
energy reduces with `\textit{y}' as expected (bandwidth decreases the activation
energy). Previously, it was shown that the increase of $\langle r_A\rangle$ converts the
charge ordered state to electronically phase-separated state~\cite{Gutierrez, Rao, Kumar1,Shankar}
and results a spontaneous metal insulator transition. Though in our case, there is no
spontaneous metal insulator transition, but the reduction of resistivity and softening of
CO state with `\textit{y}' points towards a phase coexistence scenario at larger $y$.

\begin{figure}[h!]
\includegraphics[width=0.48\textwidth]{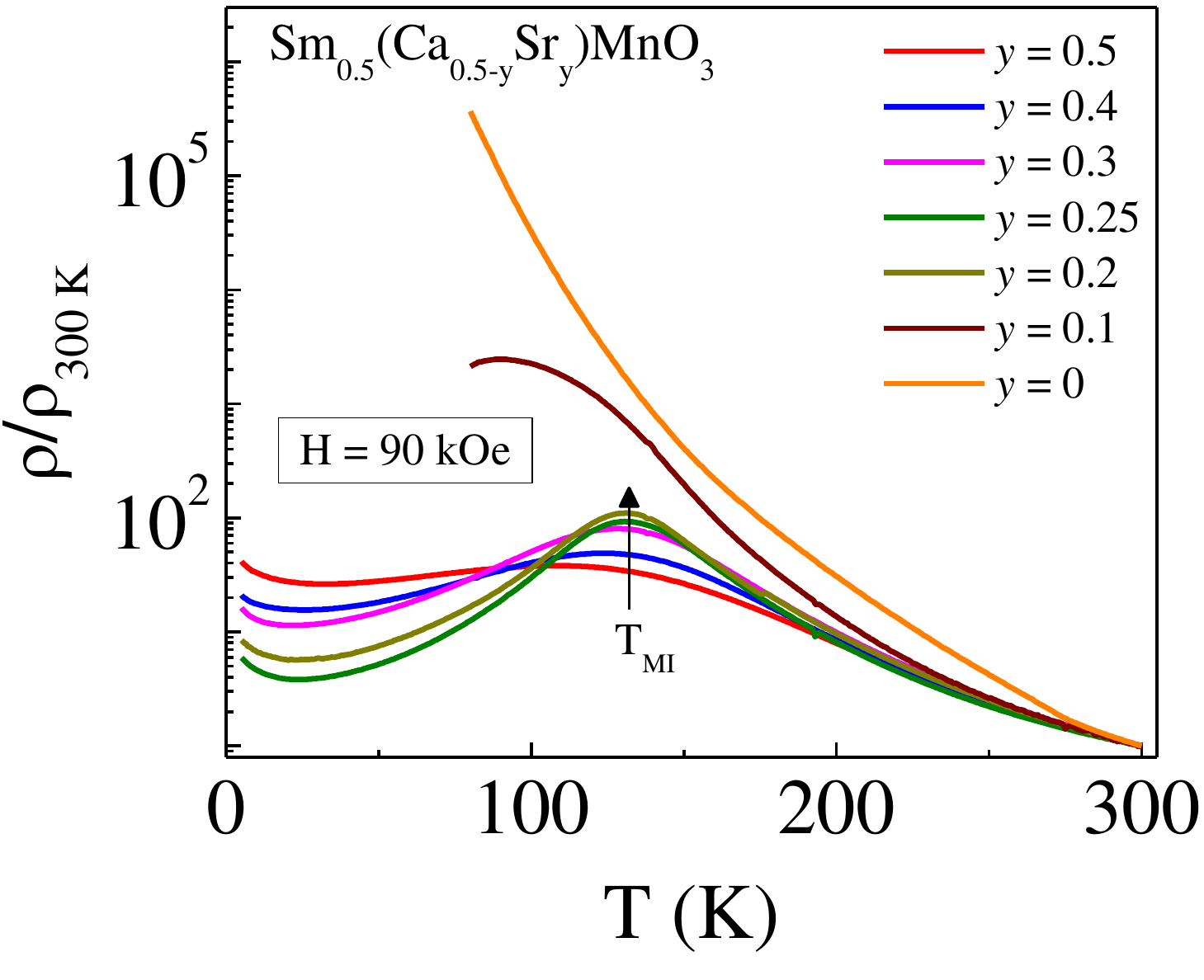}
\centering \caption[ ]
{\label{RTH} Temperature dependence of resistivity for the samples
  $Sm_{0.5}(Ca_{0.5-y}Sr_{y})MnO_3$ (\textit{y}=0, 0.1, 0.2, 0.25, 0.5) in presence of
  90 kOe external magnetic field.}
\end{figure}

To have a clear vision about this phase coexistence scenario we perform the temperature
variation of resistivity in presence of 90 kOe magnetic field (see Fig.~\ref{RTH}).
There is almost no effect of 90 kOe field on the resistivity of ${y = 0}$ sample as
expected (critical field for robust charge ordered material $Sm_{0.5}Ca_{0.5}MnO_3$ is
470 kOe at 4K). With substitution of `Sr' in place of `Ca' a huge suppression of
resistivity is observed on application of 90 kOe magnetic field and shows an insulator
to metal transition. Moreover, with increasing `Sr' concentration the curve around the
$T_{MI}$ gets broaden, which signifies the enhancement of phase coexistence. Thus, it
can be firmly said that the `Sr' doping weaken the robustness of CO state and introduces
the phase separation.

\begin{figure}[h!]
\includegraphics[width=0.48\textwidth]{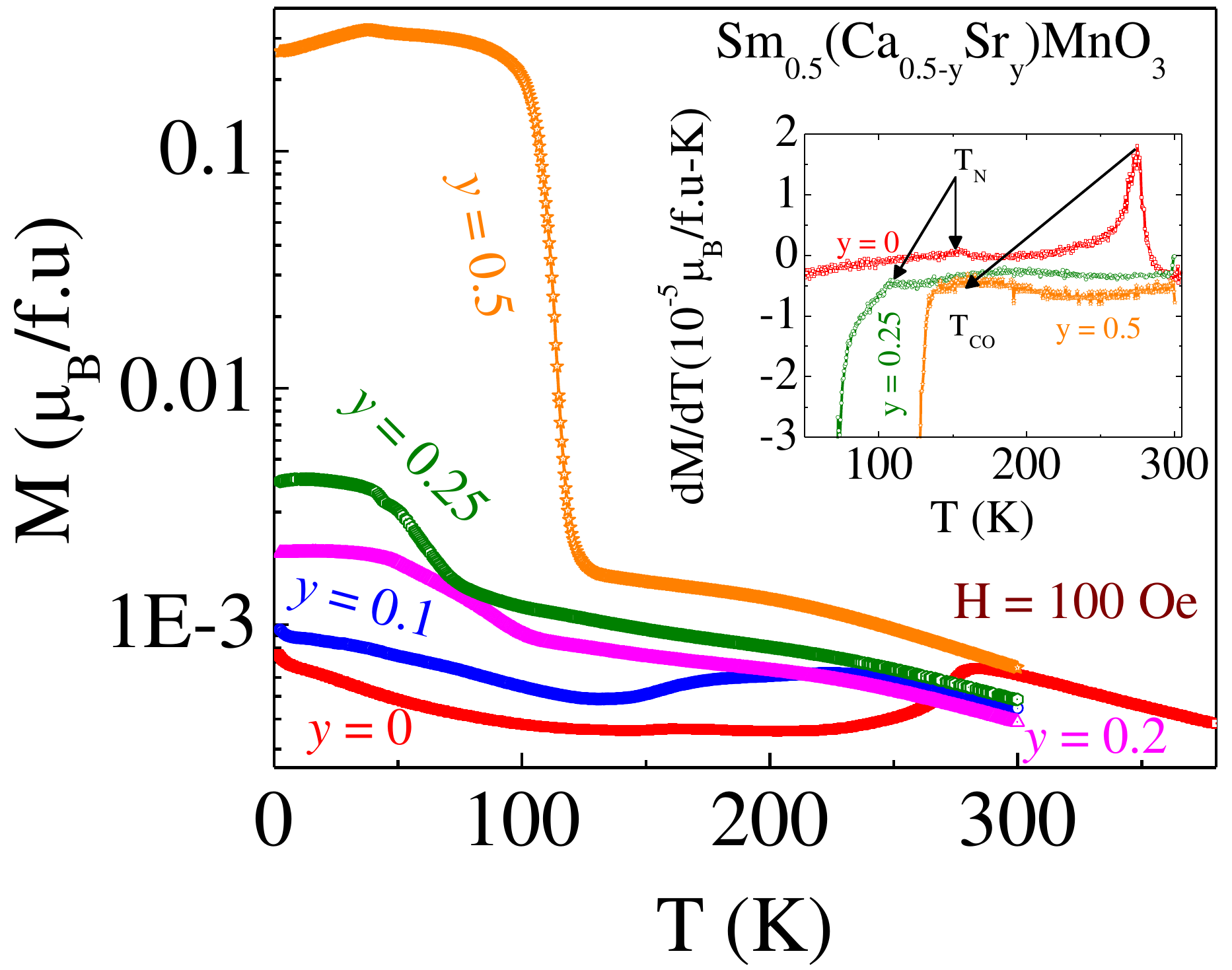}
\centering \caption[ ]
{\label{MT} Evolution of magnetization with temperature, measured in FCW protocol
in presence of 1 kOe magnetic field for the samples $Sm_{0.5}(Ca_{0.5-y}Sr_{y})MnO_3$
(\textit{y}=0, 0.1, 0.2, 0.25, 0.5). Inset shows the temperature derivative of the
corresponding magnetization of the samples.}
\end{figure}

Manganite systems being strongly correlated in nature, the phenomena of phase
separation should also be reflected in magnetization data. In this regard,
magnetization as a function of temperature [M(T)] has been measured in the field
cooled warming (FCW) protocol in presence of 100 Oe magnetic field for
(${y = 0, 0.1, 0.2, 0.25, 0.5}$). The evolution of M(T) for these samples are shown
in Fig.~\ref{MT}. At low temperature (${T < 50 K}$) with `Sr' doping the magnetization
increases. The value of magnetization increases from ${0.004\mu_B}$ to ${0.303\mu_B}$
at ${25 K}$ with increasing `Sr' concentration from ${y = 0.25}$ to ${y = 0.5}$. This
result verifies the enhancement of ferromagnetic phase fraction with `Sr' doping. At
the same time `Sr' substitution decreases the charge ordering temperature ($T_{CO}$)
as well as the anti-ferromagnetic ordering temperature (${T_N}$) (see the inset of
Fig.~\ref{MT}). For ${y = 0.5}$ composition there is no signature of $T_N$. This is
due to the existence of tri-critical point around ${100 K}$~\cite{Hiraka}.

\begin{figure}[h!]
\includegraphics[width=0.48\textwidth]{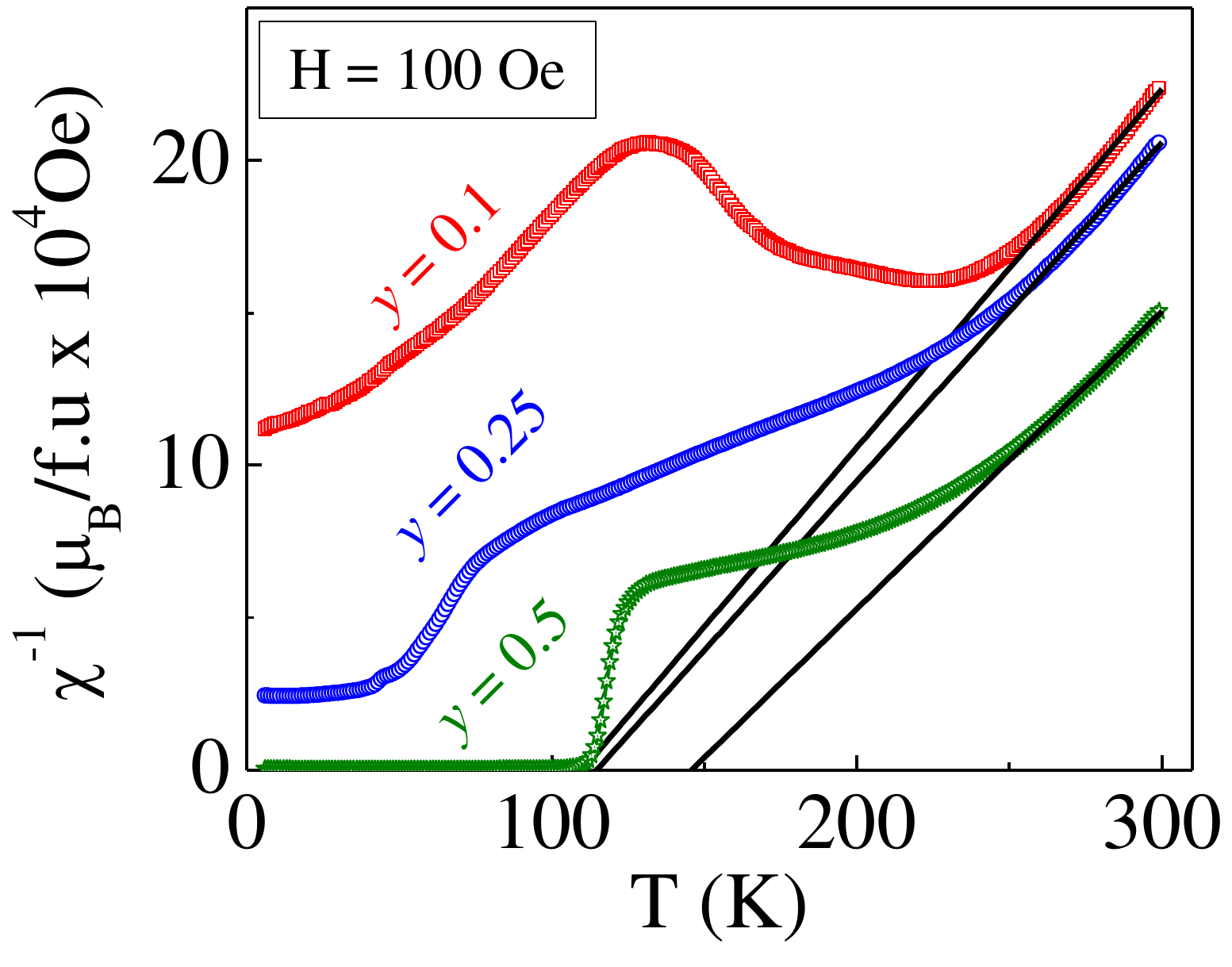}
\centering \caption[ ]
{\label{KT} Temperature dependence of inverse dc susceptibility (H/M) data, measured in
  presence of 100 Oe magnetic field for the samples $Sm_{0.5}(Ca_{0.5-y}Sr_{y})MnO_3$
  (\textit{y} = 0, 0.25, 5).}
\end{figure}

For further investigation, we analyze the high temperature (${T > 260K}$) inverse dc
susceptibility (H/M) versus temperature data at 100 Oe magnetic field with the
Curie-Weiss law $\chi = C/(T-\theta_{CW})$ where $C = \mu_{eff}^2/3 k_B $
[$\mu_{eff}$ and $\theta_{CW}$ are the effective paramagnetic moment in Bohr magneton
and paramagnetic curie temperature, respectively]. The variation of dc susceptibility
(H/M) with temperature for ${y = 0.1, 0.25, 0.5}$ samples and their corresponding
Curie-Weiss fitted data are presented in Fig.~\ref{KT}. From the fitting, paramagnetic
curie temperature comes out to be 110 K, 122 K and 145 K for the samples with
`Sr' concentrations ${y = 0.1, 0.25}$ and ${0.5}$, respectively. Furthermore, the
enhancement of the effective paramagnetic moment ($\mu_{eff} = 6.16\mu_B$ for
${y = 0.1}$ to $\mu_{eff} = 6.76\mu_B$ for ${y = 0.5}$) has also been observed. This
increase of $\theta_{CW}$ and $\mu_{eff}$ clearly indicates the enhancement of
ferromagnetic interactions with `Sr' doping. Here another point needs to mention is
that the values of the effective moments are larger than the theoretical calculated
value of ${4.42\mu_B}$. It indicates the presence of ferromagnetic clusters in the
high temperature region (${T > 260K}$). These clusters behaves as an individual
paramagnetic entity which contains more than one Mn ions~\cite{Banikrsc}. This is why
$\mu_{eff}$ value [= ${4.32\mu_B}$] for ${y = 0}$ sample (without any ferromagnetic
clusters) is close to the theoretical expected value. These systematic result further
implies that `Sr' doping induces the ferromagnetic clusters in the host antiferromagnetic
phase.

\begin{figure}[h!]
\includegraphics[width=0.48\textwidth]{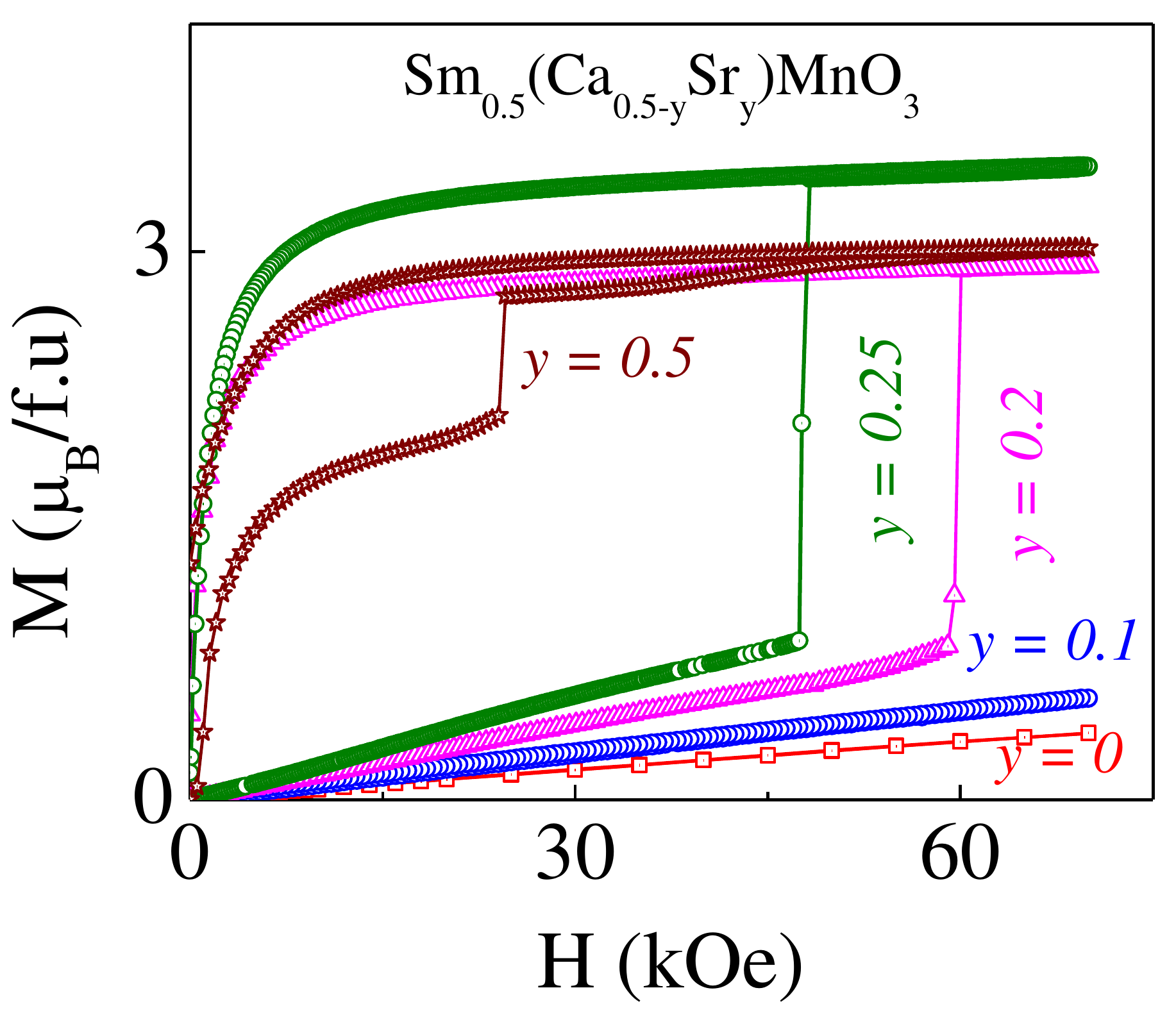}
\centering \caption[ ]
{\label{MH} Isothermal magnetization of the samples $Sm_{0.5}(Ca_{0.5-y}Sr_{y})MnO_3$
(\textit{y} = 0, 0.1, 0.2, 0.25, 5) at 2 K.}
\end{figure}

\subsection{Metamagnetic transition}

To see the effect of magnetic field in these phase separated state, we measure the
field dependence of magnetization at ${2 K}$ for these samples as presented in
Fig.~\ref{MH}. For the samples without any `Sr' concentrations (i.e. ${y = 0}$) and
with small `Sr' concentrations (i.e. ${y = 0.1}$) we find a linear increase of
magnetization with magnetic field up to the 70 kOe field due to the strong CO-AFM phase
in parent compound SCMO. With further increase of `Sr' concentration i.e. to
${y = 0.2}$, a sharp metamagnetic transition is observed at 55 kOe (the magnetization
increases from ${0.75\mu_B}$ to ${3\mu_B}$). So at this point the system converts
completely from a CO antiferromagnetic (CO-AFM) phase to a ferromagnetic (FM) phase.
The descending branch of the M-H curve remains almost flat down to 10 kOe field which
indicates the irreversible nature of the field induced CO-AFM to FM transformation.
With further decreasing of the field from 10 kOe we find that the magnetization
rapidly decreases. On the other hand, with further increase of `Sr' concentrations
requirement of critical field for the metamagnetic transition decreases to 47 kOe for
${y = 0.25}$ and to 20 kOe for ${y = 0.5}$. Although in ${y = 0.5}$, the initial increase
of magnetization is like a soft ferromagnet, which indicates the dominance of ferromagnetic
phases as compared to the antiferromagnetic phases in the sample. Previously, the same
kind of sharp metamagnetic transition has been observed in Mn site doped CO manganites,
for instance in $Pr_{0.5}Ca_{0.5}Mn_{1-x}M_xO_3$. According to Raveau et al.~\cite{Mahendiran}
the occurrence of this step like behavior in Mn site doped manganites is because of
the presence of the short range ordered ferromagnetic regions in the CO-AFM region.
Here, the existence of ferromagnetic clusters and their growth with `Sr' doping is
also observed from the analysis of dc susceptibility data. These ferromagnetic clusters
play the role of nucleation centers in the metamagnetic transitions (from CO-AFM to
FM phase) and sharpness of steps indicates the jerky growth of these FM clusters. The
downward shifting of the critical fields with increasing `Sr' concentrations is possibly 
because of the increasing number of ferromagnetic clusters.

\begin{figure}[h!]
\includegraphics[width=0.48\textwidth]{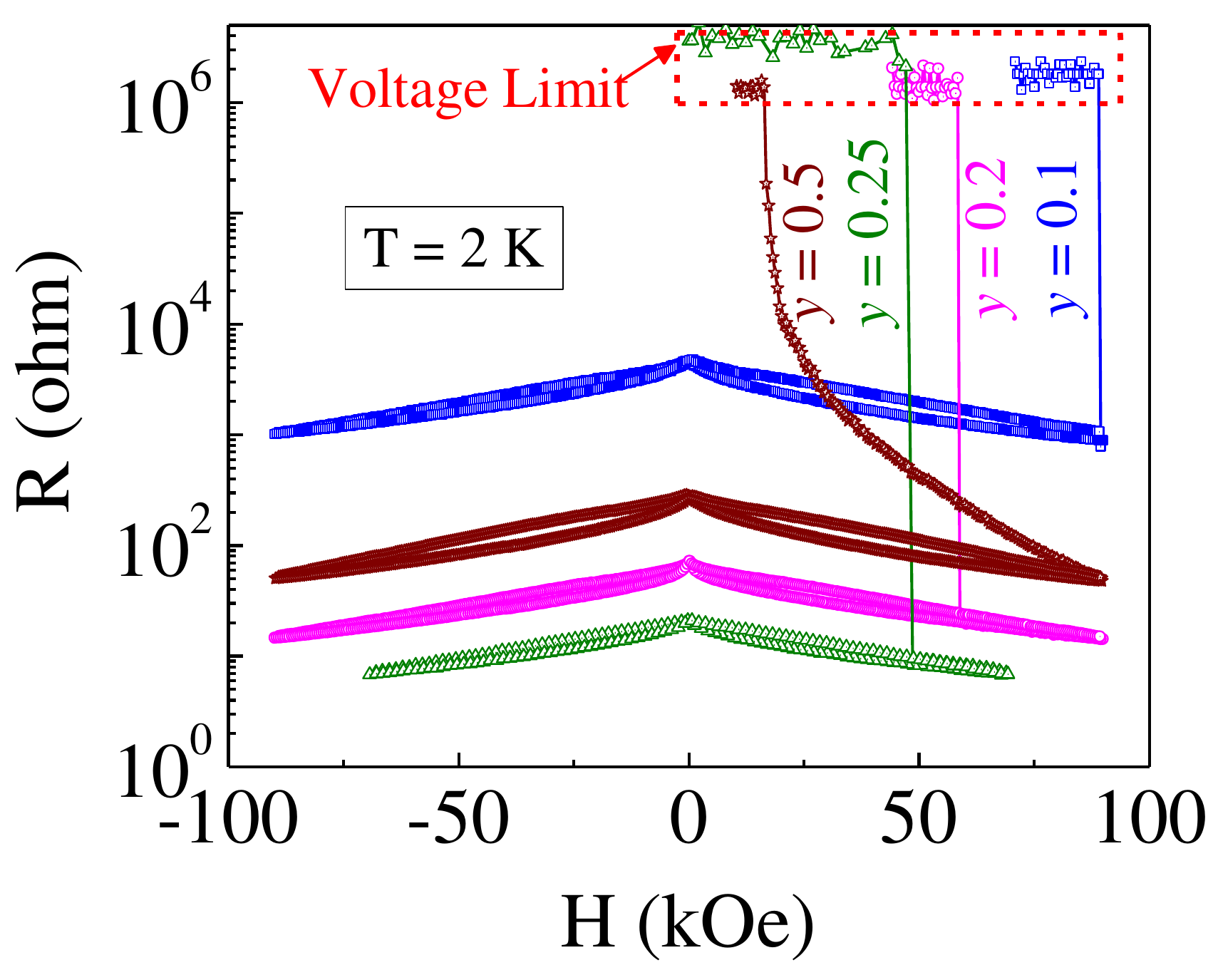}
\centering \caption[ ]
{\label{RH} Magnetic field dependence of resistance of the samples
  $Sm_{0.5}(Ca_{0.5-y}Sr_{y})MnO_3$ at 2 K.}
\end{figure}

Next, we measure the resistance with variation of magnetic fields at 2 K for our samples
to correlate the shifting of the critical field as seen in magnetization data. Here in
the sample ${y = 0.1}$, metamagnetic transition is observed
at 89.7 kOe which was not visible in magnetization because of the instrumental limitation.
In ${y = 0}$ sample there is almost no effect of magnetic field (not shown in the figure).
In the subsequent `Sr' doped samples i.e. for ${y = 0.2}$ metamagnetic transition occurs
at 58.8 kOe [and 48.5 kOe for ${y = 0.25}$]. The slight higher values of the critical field
determined from R-H compared with M-H is possibly due to different average sweep rate
of the field. For example for ${y = 0.25}$ sample, in M-H  sweep rate was 100 Oe/sec
and in R-H it was 24 Oe/Sec. The dependence of critical field on sweep rate is a signature
of martensitic like transition as smaller sweep rate assist the progressive accommodation
of the martensitic strain which push the instability towards higher magnetic field as
discussed in our earlier work~\cite{Sanjibnpg}.

\subsection{Theoretical study}

In order to further explain the experimental results we perform spin-fermion Monte Carlo
calculations. Our prime motive is to analyze the role of `Sr' disorder on the metamagnetic
transition. In SCMO, 'Sr' doping is simulated by adding quenched disorder to a well
studied model Hamiltonian for manganites in the large Hund's coupling limit
$(J_H\rightarrow\propto)$~\cite{Dagotto1,Pradhanprl}. Our model Hamiltonian:

\begin{eqnarray*}
H &=& -\sum_{\langle ij \rangle \sigma}^{\alpha \beta}
t_{\alpha \beta}^{ij}
 c^{\dagger}_{i \alpha \sigma} c^{~}_{j \beta \sigma}
~ - J_H\sum_i {\bf S}_i.{\mbox {\boldmath $\sigma$}}_i \cr
&&
~~+ J \sum_{\langle ij \rangle}
{\bf S}_i.{\bf S}_j
 - \lambda \sum_i {\bf Q}_i.{\mbox {\boldmath $\tau$}}_i
+ {K \over 2} \sum_i {\bf Q}_i^2
\end{eqnarray*}

\noindent

where $t_{\alpha\beta}$ is the hopping amplitude between nearest neighbor $e_g$
electrons with orbitals ($\alpha$, $\beta$) and $\sigma\,(=\uparrow,\downarrow)$.
$\alpha$ (and $\beta$) denotes $d_{x^2 - y^2}$ and $d_{3z^2 - r^2}$ Mn-orbitals.
$J_H$ (Hund's coupling) is between Mn $t_{2g}$ spin ${\bf S_i}$ and $e_g$ electron
spin {\boldmath $\sigma_i$} at site $i$, whereas J is the superexchange
interaction between neighboring Mn $t_{2g}$ spins. $\lambda$ is the electron-phonon
interaction between $e_g$ electrons and the Jahn-Teller phonon
modes $(\textbf{Q}_i)$ in the adiabatic limit. We treat $(\textbf{S}_i)$  and
$(\textbf{Q}_i)$ as classical~\cite{Dagotto2} variables. $K$ (stiffness of
Jahn-Teller modes) and $|{\bf S}_i|$ are set to be 1. For more details please
see Ref.\citenum{Dagotto1}

We incorporated the effect of 'Sr' disorder by adding $\sum {\epsilon_i n_i}$ term to
the Hamiltonian. One generally add $\sum_i \epsilon_i n_i$ to the Hamiltonian such that
average $\overline{\epsilon}_i = 0$ to model A-site disorder in manganites with two A-type
elements~\cite{Tokura,Anamitra} [example: for $Sm_{0.5}Ca_{0.5}MnO_3$ (SCMO) or
$Sm_{0.5}Sr_{0.5}MnO_3$ (SSMO) like samples]. This is done by choosing $\epsilon_i$ at
each site from 
$P(\epsilon_i) = {1 \over 2} \delta(\epsilon_i - \Delta_H) + {1 \over 2} \delta(\epsilon_i + \Delta_H)$
distribution. In $Sm_{0.5}Ca_{0.25}Sr_{0.25}MnO_3$ (SCSMO), with three A-type elements,
${Sr^{2+}}$ ions (with larger ionic radius than $Sm^{3+}$ and $Ca^{2+}$ ions) occupy
one fourth of the A-sites randomly. So by neglecting the ionic mismatch between Sm and
Ca elements we model Sr and Sm-Ca disorder by adding $\sum_i \epsilon_i n_i$ at each Mn
site picked from the distribution
$P(\epsilon_i) = {1 \over 4} \delta(\epsilon_i - \Delta_Q) + {3 \over 4} \delta(\epsilon_i + \Delta_Q)$~\cite{Sanjibnpg,Sanjibarxiv}.
Both $\Delta_H$ and $\Delta_Q$ are quenched disorder potentials. $\Delta_H$  (H stand for half)
and $\Delta_Q$  (Q stand for quartern) are binary disorder with ratio 50-50 and 25-75,
respectively. In an external magnetic field we add a Zeeman coupling term
$-\sum_i {\bf h}\cdot{\bf S}_i$ in our model Hamiltonian. We measure all the parameters in terms
of the hopping energy $t$. The estimated value of $t$ in manganites is
0.2 eV~\cite{Dagotto1}.

We use an exact diagonalisation scheme for the $e_g$ electrons using different
background $t_{2g}$ spins and phonon mode $(\textbf{Q})$ configurations. Background
configurations were chosen using travelling cluster approximation (TCA) based spin-fermion
Monte Carlo technique to access the large system size $(24 \times 24$ lattice)~\cite{Kumar,Pradhanprl}.
In different magnetic field our measured magnetization is thermally averaged over ten
different disorder samples in addition to the thermal averages during the Monte Carlo
sweeps. Over 10,000 Monte Carlo sweeps were performed to thermalize the system.

\begin{figure}[h!]
\includegraphics[width=0.48\textwidth]{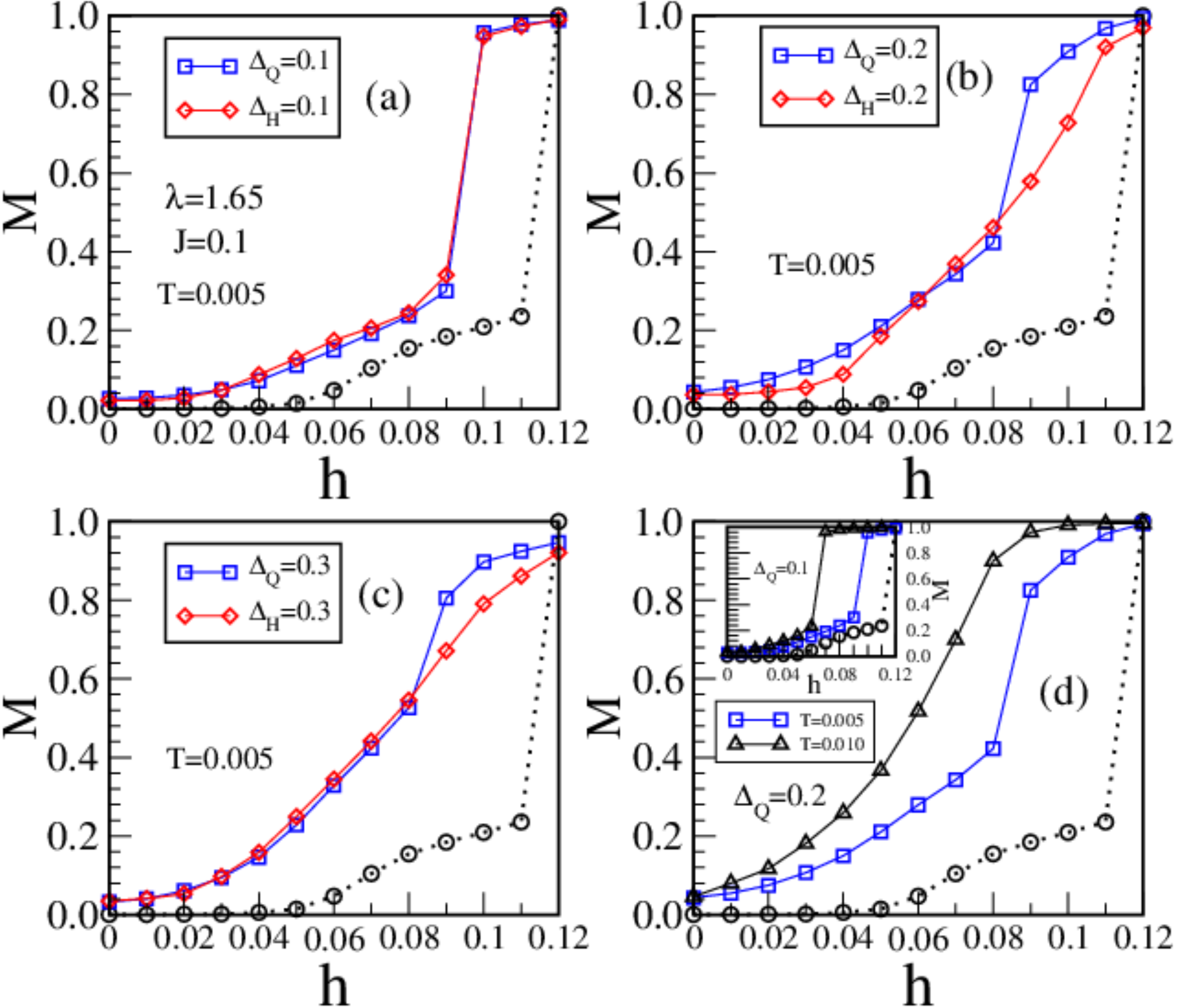}
\caption{ Magnetization vs. field using $\lambda$= 1.65 and $J$ = 0.1. 
  (a)--(c) $M-h$ phase diagrams using different $\Delta$ (both $\Delta_Q$ and $\Delta_H$)
  values at temperature T = 0.005. Dotted line (circle symbols) in (a) [also in (c)--(d)]
  is for clean system ($\Delta$ = 0). (d) $M-h$ phase diagram for $\Delta_Q$ =0.2 for
  T = 0.005 and T = 0.01. Inset of (d) shows the same using $\Delta_Q$ =0.1. }
\label{th-fig1} 
\end{figure}

First we start our calculations using $\lambda$ = 1.65 and $J$ = 0.1 that reproduces the
correct magnetic phase (CE-CO-OO-I phase) at low temperatures for electron density
$n$ = $1-x$ = 0.5~\cite{Pradhanprb}. At T = 0.005 the magnetization (M) remains very small
even for $\Delta_Q$ = 0.3~\cite{Sanjibnpg}. In Figs.~\ref{th-fig1}(a)--(d) we show the
magnetization vs. field ($h$) curve for the clean system (without any disorder) using dotted lines.
The metamagnetic transition is at $h$ = 0.11. For $\Delta_Q$ = 0.1 the metamagnetic transition
remains sharp, but the critical field $h_{cr}$ decreases to 0.09 as shown in Fig.~\ref{th-fig1}(a).
For $\Delta_H$ = 0.1, the critical field for the magnetic transition remains the same to that of
$\Delta_Q$ = 0.1. So, for $\Delta_Q$ = $\Delta_H$ = 0.1 the system behaves more or less
like clean system. Also the inset of Fig.~\ref{th-fig1}(d) shows that the magnetic
transition for $\Delta_Q$ = 0.1 remains sharp even for higher temperature (T = 0.01) unlike
SCSMO experiments~\cite{Sanjibnpg}. Therefore we believe that the disorder strength is larger
that $\Delta_Q$ = 0.1 in SCSMO. For $\Delta_H$ = 0.2 and 0.3 the magnetic transition is
continuous with the field and qualitatively agrees with previous work~\cite{Anamitra}. But the
correct way to model 'Sr' disorder in SCSMO is to add disorder that is $\Delta_Q$ type. It
is apparently clear that the critical field for metamagnetic transition decreases to $h$ = 0.08
for $\Delta_H$ = 0.2 and 0.3.  For $\Delta_Q$ = 0.2, the sharpness of magnetic transition
vanishes at T = 0.01, which qualitatively matches with the trend of SCSMO
experimental data~\cite{Sanjibnpg}. 

\begin{figure}[h!]
\includegraphics[width=0.48\textwidth]{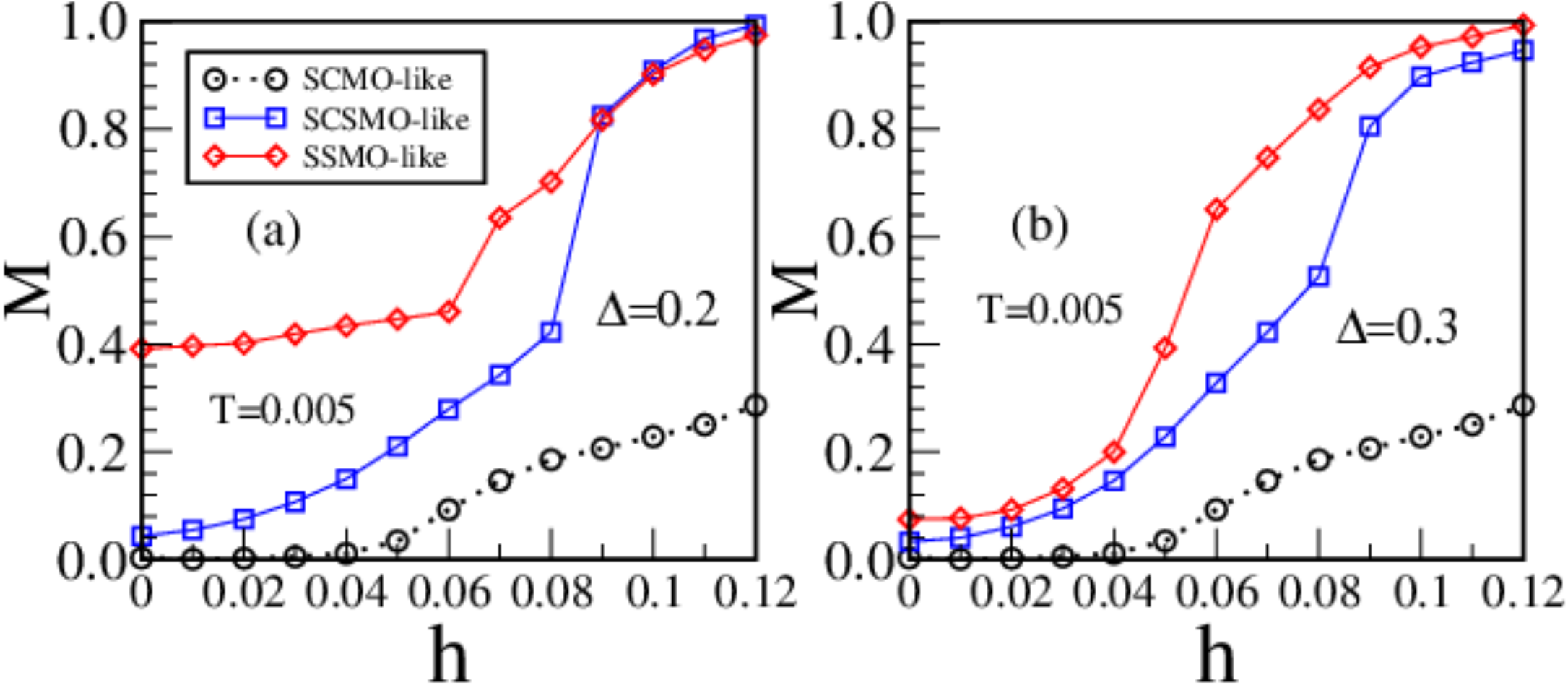}
\caption{ Magnetization	vs. field ($M-h$) for three sets of parameters
  (mimicking SCMO, SCSMO and SSMO samples). (a) $\Delta$ =0.2 and (b)  $\Delta$ =0.3.
  Dashed lines in both (a) and (b) are for $\Delta$ = 0. Please see the text for details.}
\label{th-fig2}
\end{figure}

Next, we move to analyze the M-h curve for the series (SCMO-like, SCSMO-like and SSMO-like)
of materials as prepared in the experiments. We know that SSMO (SCMO) has larger (smaller)
bandwidth than SCSMO. We incorporate the bandwidth variation by changing $\lambda$ (and $J$)
values in our model Hamiltonian. Smaller $\lambda$ (and $J$) implies larger bandwidth or
vice versa. For clarity we treat SCMO as a clean system (due to the small mismatch between
Sm and Ca ionic radii) and use $\lambda$= 1.73, $J$ = 0.105 for SCMO-like materials. For
SSMO-like materials we set $\Delta_H$ = 0.2,  $\lambda$= 1.57, $J$ = 0.095. For SCSMO we
choose $\Delta_Q$ = 0.2,  $\lambda$= 1.65, $J$ = 0.1 as discussed above. Fig.~\ref{th-fig2}(a)
shows that our results qualitatively agree with the experiments. We denote $\Delta$ = 0.2 (0.3)
[i.e. $\Delta_Q$ = 0.2 (0.3) for SCSMO-like material and $\Delta_H$ = 0.2 (0.3) for SSMO like
  material] for brevity. Magnetization of SCMO-like material remains small even for $h = 0.12$.
For SCSMO-like material the metamagnetic transition is at critical field $h_{cr} = 0.08$.
For SSMO we find sizeable magnetization even at $h = 0.0$ and the metamagnetic transition
is at $h = 0.06$ (smaller than that of SCSMO). For clarity and completeness
we also plot the two curves using $\Delta_Q$= 0.3 and $\Delta_H$= 0.3 (along
with the clean SCMO) in Fig.~\ref{th-fig2}(b). The trend of M-H curves from SCMO-like to
SSMO-like materials remains qualitatively same for both set of parameters.

\begin{figure}[h!]
\includegraphics[width=0.48\textwidth]{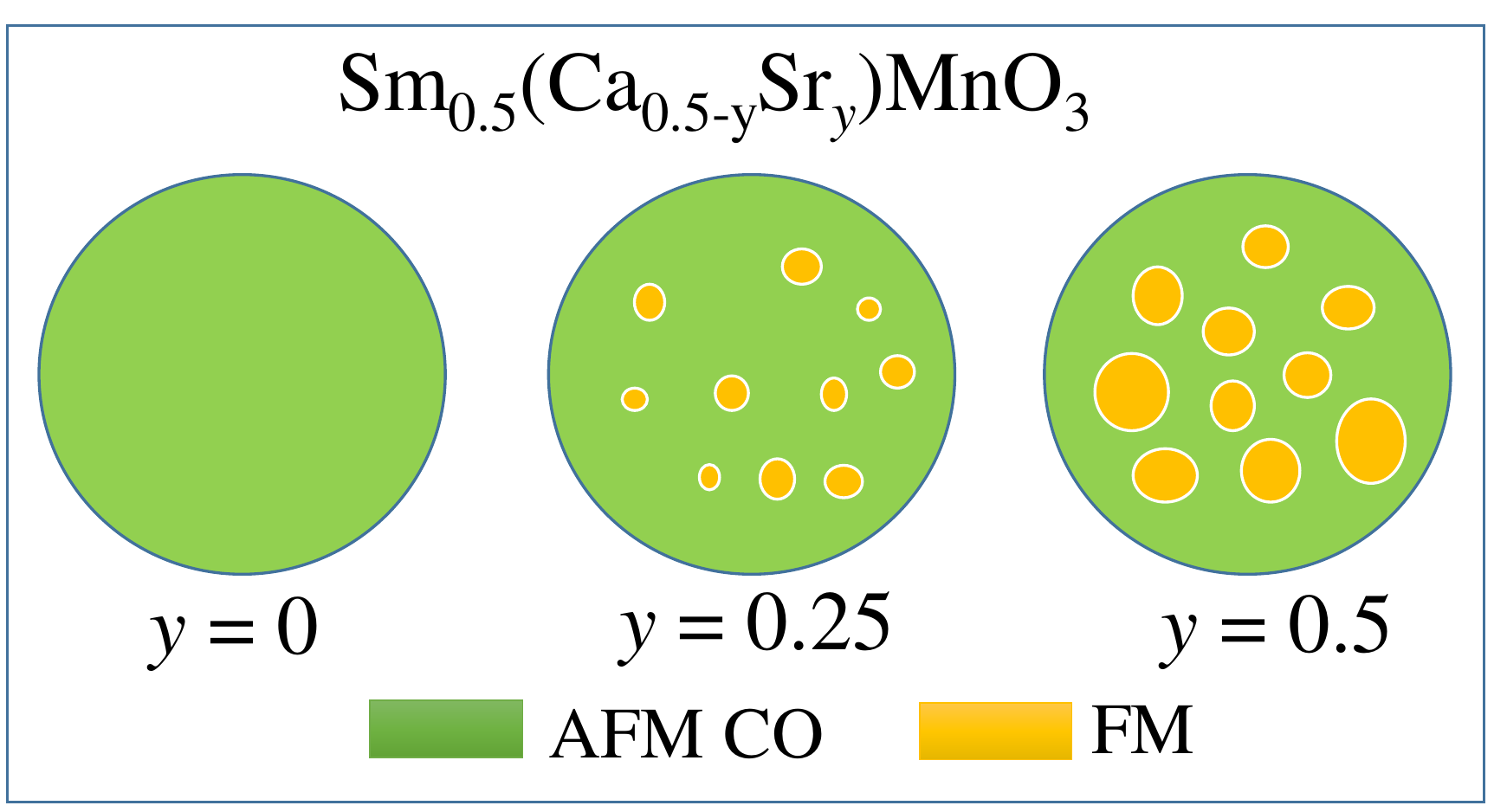}
\centering \caption[ ]
{\label{schematic} A schematic picture to describe the reduction of critical
  magnetic field for the metamagnetic transition from $y$ = 0 to $y$ =0.5.}
\end{figure}

So the experimental scenario can be explained by a simple phenomenological picture
obtained from our experimental and theoretical study (see  Fig.~\ref{schematic}). According
to this picture `Sr' doping (in place of `Ca') in SCMO induces ferromagnetic clusters. The
area and number of ferromagnetic cluster increases gradually with `Sr' doping and at
${y = 0.5}$ area of each ferromagnetic cluster phase is maximum. These clusters act as the
nucleation center that give rise to martensitic like transformation and convert the CO-AFM
phase to a FM phase. With increasing the bandwidth (by increasing $y$) the phase separation
increases, i.e. the strength as well as number of these nucleation center increases and as
a result the critical field decreases with $y$.

\section{Conclusions}

In summary, we have investigated the metamagnetic properties of the
$Sm_{0.5}(Ca_{0.5-y}Sr_{y})MnO_3$ compounds through isothermal magnetization and
resistivity measurements. The presence of ultra-sharp jump at low temperature
($T < 5 K$) in both isothermal resistivity and magnetization is due to the 
strong spin and charge coupling in the systems. The sweep rate dependence of the critical
field ($H_{cr}$) indicates the martensitic scenario. `Sr' doping (in place of `Ca') in
SCMO induces ferromagnetic clusters and the volume of these ferromagnetic clusters 
increases in the CO-AFM materials with average A-site ionic radius $\langle r_A\rangle$ as
perceived from the magnetotransport and magnetization data. Our model Hamiltonian
calculations also confirm this scenario. These ferromagnetic clusters act as the
nucleation center in the CO-AFM background for the burst like growth from CO-AFM to
FM phase at critical field. The FM fraction increases with $\langle r_A\rangle$ and
as result the $H_{cr}$ decreases with 'Sr' doping.

\section{Acknowledgements}
The work was supported by Department of Atomic Energy (DAE), Govt. of India.


\begin{thebibliography}{abc}
\bibitem{Mahendiran} R. Mahendiran, A. Maignan, S. Hebert, C. Martin, M. Hervieu,
B. Raveau, J.F. Mitchell, and P. Schiffer, Phys. Rev. Lett. \textbf{89}, 286602 (2002).

\bibitem{Autret} C. Autret, A. Maignan, C. Martin, M. Hervieu, V. Hardy, S. Hebert,
  and B. Raveau, Appl. Phys. Lett. \textbf{82}, 4746 (2003).

\bibitem{Fisher} L.M. Fisher, A. V. Kalinov, I.F. Voloshin, N.A. Babushkina, D.I. Khomskii,
  Y. Zhang, and T.T.M. Palstra, Phys. Rev. B \textbf{70}, 212411 (2004).

\bibitem{Hardy1} V. Hardy, S. Majumdar, S.J. Crowe, M.R. Lees, D. Mck Paul, L. Herve,
  A. Maignan, S. Hebert, C. Martin, C. Yaicle, M. Hervieu, and B. Raveau,
  Phys. Rev. B \textbf{69}, 20407 (2004).

\bibitem{KrishnaM} J. Krishna Murthy, K.D. Chandrasekhar, H.C. Wu, H.D. Yang, J.Y. Lin,
  and A. Venimadhav, EPL (Europhysics Lett. \textbf{108}, 27013 (2014).

\bibitem{Ouyang} Z.W. Ouyang, H. Nojiri, and S. Yoshii, Phys. Rev. B \textbf{78}, 104404 (2008).

\bibitem{Roy} S.B. Roy, M.K. Chattopadhyay, P. Chaddah, and A.K. Nigam,
  Phys. Rev. B \textbf{71}, 174413 (2005).

\bibitem{Velez} S. Velez, J.M. Hernandez, A. Fernandez, F. Macie, C. Magen, P.A. Algarabel,
  J. Tejada, and E.M. Chudnovsky, Phys. Rev. B \textbf{81}, 64437 (2010).

\bibitem{Choi} Y.J. Choi, C.L. Zhang, N. Lee, and S.-W. Cheong,
  Phys. Rev. Lett. \textbf{105}, 97201 (2010).

\bibitem{Danjoh} S. Danjoh, J.-S. Jung, H. Nakamura, Y. Wakabayashi, and T. Kimura,
  Phys. Rev. B \textbf{80}, 180408 (2009).

\bibitem{Hardy2} V. Hardy, A. Maignan, S. Hebert, C. Yaicle, C. Martin, M. Hervieu,
  M.R. Lees, G. Rowlands, D.M.K. Paul, and B. Raveau, Phys. Rev. B \textbf{68}, 220402 (2003).

\bibitem{Wu} T. Wu and J.F. Mitchell, Phys. Rev. B \textbf{69}, 100405 (2004).

\bibitem{Liao} D. Liao, Y. Sun, R. Yang, Q. Li, and Z. Cheng,
  Phys. Rev. B \textbf{74}, 174434 (2006).

\bibitem{Cao} G. Cao, J. Zhang, S. Cao, C. Jing, and X. Shen,
  Phys. Rev. B \textbf{71}, 174414 (2005).

\bibitem{Bordel} C. Bordel, J. Juraszek, D.W. Cooke, C. Baldasseroni, S. Mankovsky, J. Minaer,
  H. Ebert, S. Moyerman, E.E. Fullerton, and F. Hellman, Phys. Rev. Lett. \textbf{109}, 117201 (2012).

\bibitem{Tsui} Y.K. Tsui, C.A. Burns, J. Snyder, and P. Schiffer, Phys. Rev. Lett. \textbf{82}, 3532 (1999).

\bibitem{Flint} R. Flint, H.-T. Yi, P. Chandra, S.-W. Cheong, and V. Kiryukhin, Phys. Rev. B \textbf{81}, 92402 (2010).

\bibitem{Hardy3} V. Hardy, S. Hebert, A. Maignan, C. Martin, M. Hervieu, and B. Raveau,
  J. Magn. Magn. Mater. \textbf{264}, 183 (2003).

\bibitem{Tokura} Y. Tokura, Reports Prog. Phys. \textbf{69}, 797 (2006).

\bibitem{Moritomo} Y. Moritomo, H. Kuwahara, Y. Tomioka, and Y. Tokura, Phys. Rev. B \textbf{55}, 7549 (1997).

\bibitem{Mathieu} R. Mathieu, M. Uchida, Y. Kaneko, J.P. He, X.Z. Yu, R. Kumai, T. Arima,
  Y. Tomioka, A. Asamitsu, Y. Matsui, and Y. Tokura, Phys. Rev. B \textbf{74}, 20404 (2006).

\bibitem{Banik} S. Banik, K. Das, and I. Das, J. Magn. Magn. Mater. \textbf{403}, 36 (2016).


\bibitem{Gutierrez} D. Gutierrez, G. Radaelli, F. Sanchez, R. Bertacco, and J. Fontcuberta, Phys. Rev. B \textbf{89}, 075107 (2014).

\bibitem{Rao} C. N. R. Rao, Anthony Arulraj, P. N. Santosh, and A. K. Cheetham, Chem. Mater. \textbf{10}, 2714-2722 (1998).

\bibitem{Kumar1} N. Kumar, and C. N. R. Rao, Journal of Solid State Chemistry \textbf{129}, 363-366 (1997)

\bibitem{Shankar} U. Shankar and A. K. Singh, J. Phys. Chem. C \textbf{119}, 28620−28630 (2015).


\bibitem{Hiraka} Y. Tomioka, H. Hiraka, Y. Endoh, and Y. Tokura, Phys. Rev. B \textbf{74}, 104420 (2006).

\bibitem{Banikrsc} S. Banik, K. Das and I. Das, RSC Adv. \textbf{7}, 16575 (2017).

\bibitem{Sanjibnpg} S. Banik et al. NPG Asia Materials {\bf 10} , 923 (2018).
  
\bibitem{Dagotto1} E. Dagotto, T. Hotta, and A. Moreo, Phys. Rep. {\bf 344}, 1 (2001).

\bibitem{Pradhanprl} K. Pradhan, A. Mukherjee, and P. Majumdar, Phys. Rev. Lett. {\bf 99}, 147206 (2007).

\bibitem{Dagotto2} E. Dagotto, S. Yunoki, A. L. Malvezzi, A. Moreo, J. Hu, S. Capponi, D. Poilblanc,
  and N. Furukawa, Phys. Rev. B {\bf 58}, 6414 (1998).

  
\bibitem{Anamitra} A. Mukherjee, K. Pradhan and P. Majumdar, Europhys. Lett. {\bf 86}, 27003 (2009).

\bibitem{Sanjibarxiv} S. Banik, K. Das, K. Pradhan and I. Das, arxiv: 1902.04377.

\bibitem{Kumar} S. Kumar and P. Majumdar, Eur. Phys. J. {\bf B 50}, 571 (2006).

\bibitem{Pradhanprb}
K. Pradhan $\&$ S. Yunoki, Phys. Rev. B {\bf 96}, 214416 (2017).

\end{thebibliography}
\end{document}